\documentclass[conference]{IEEEtran}
\usepackage{graphics}
\usepackage{graphicx}
\usepackage{epsfig,amssymb}
\usepackage{epstopdf}
\usepackage{amsmath}
\usepackage{breqn}
\usepackage{times}
\usepackage{mathrsfs}
\usepackage{array}
\usepackage{amsmath, latexsym, amsfonts, amssymb}
\usepackage[mathscr]{eucal}
\usepackage{array, tabularx}
\usepackage[table]{xcolor}
\usepackage{multirow}
\usepackage{epsfig}

\usepackage{cite}
\usepackage[numbers,sort & compress]{natbib}

\usepackage{tikz}
\usepackage{url}
\usepackage{mathtools}
\usepackage{psfragx}
\usepackage{xcolor}
\usepackage{bbm}
\usepackage{authblk}

\newcommand{\brparen}[1]{\left\{#1\right\}}

\DeclareMathOperator*{\argmax}{arg\,max}

\IEEEoverridecommandlockouts

\begin{document}

\title{Over-the-Air Implementation of Uplink NOMA }

%
%
%
%
%

\author[$ \dag $]{Samith Abeywickrama}
\author[$ \dag $]{Lei Liu}
\author[$ \ddag $]{Yuhao Chi}
\author[$ \dag $]{Chau Yuen}
\affil[$ \dag $]{Singapore University of Technology and Design,  Singapore}
\affil[$ \ddag $]{Nanyang Technological University, Singapore}
\affil[$ \dag $ ]{Email: \textit { \{abeywickrama$ \_ $samith,lei$ \_ $liu,yuenchau\}@sutd.edu.sg}}
\affil[$ \ddag $ ]{Email: \textit { \{chiyuhao1990\}@163.com} 
\thanks{This work is supported by A*STAR (Agency for Science, Technology and Research) SERC project, under the grant no. 1420200043.}
}

\date{}
\bibliographystyle{ieeetr}
\maketitle
\vspace{-0.5cm}

\begin{abstract}

Though the concept of non-orthogonal multiple access (NOMA) was proposed several years ago, the performance of uplink NOMA has only been verified in theory, but not in practice. This paper presents an over-the-air implementation of a uplink NOMA system, while providing solutions to most common practical problems, \textit{i.e.}, carrier frequency offset (CFO) synchronization, time synchronization, and channel estimation.
The implemented CFO synchronization method adopts the primary synchronization signal (PSS) of LTE. Also, we design a novel preamble for each uplink user, and it is appended to every frame before it is transmitted through the air. This preamble will be used for time synchronization and channel estimation at the BS. Also, a low-complexity, iterative linear minimum mean squared error (LMMSE) detector has been implemented for multi-user decoding. The paper also validates the proposed architecture numerically, as well as experimentally.
\end{abstract}

\section{Introduction}

Non-orthogonal multiple access (NOMA) that allows more than one user to share the same sub-carrier when users are multiplexed in the power domain, has been looked upon as a candidate for further cellular enhancements toward the 5th generation (5G) mobile communications system \cite{noma_ref_1,noma_ref_2}. Multiple users are superimposed with different power gains and separated via a Successive Interference Cancellation (SIC)  at the receiver. However, the performance of uplink NOMA has only been verified in theory, but not in practice. To this end, this paper presents an over-the-air implementation of a uplink NOMA system, while providing solutions to most important practical problems.

To the best of the authors' knowledge, most of the existing over-the-air implementations of NOMA focus on downlink transmission \cite{noma_impl_1,noma_impl_2,noma_impl_3}. For the downlink NOMA, the superimposed signals transmitted by the base station (BS) are always synchronous, as the BS controls transmission for all users \cite{noma_chall_1}. 
Being different to the downlink NOMA, time synchronization is challenging for the uplink NOMA due to the fact that the mobile communications channel is usually dynamic in nature, and users are spatially distributed \cite{noma_chall_1}. Therefore, the symbols transmitted by the superposition-coded users are time-misaligned at the BS, and it causes performance degradation of uplink NOMA. Because of this reason, uplink user equipments (UEs) should arrange their timing advance. This means, UEs start transmitting symbol sequence in such a way that all signals arrive to the BS simultaneously.

Carrier frequency offset (CFO) can be considered as one of the most common impairments found in a practical communication system. CFO occures when the frequency is not stable in the transmitter and the receiver oscillators, and also due to the Doppler effect as the transmitter or the receiver is moving \cite{cfo}. This CFO leads to a large amount of multiple access interference in the uplink NOMA system, since CFO misaligns the sub-carriers in the frequency domain. Therefore, CFO estimation and compensation becomes necessary. 

When it comes to the channel estimation of uplink NOMA, base station (BS) has to estimate the channel parameters of all the uplink users when more than one user share the same time, frequency, and
spreading code. For the downlink NOMA, channel estimation is a trivial task, since UE only has to estimate the channel between the UE and the BS. However, by exploiting the channel reciprocity,
BS can determine the uplink channel when UE estimates the downlink channel and transmit it back to the BS through a feedback channel. This method is mainly applicable for time division duplex (TDD) systems that use the same frequency for the uplink and the downlink. Also, using channel reciprocity for channel estimation leads to many practical difficulties, due to the non-symmetric characteristics of the RF front-end circuitry at the receiver and the transmitter \cite{our_tsp}. In \cite{noma_ce_ref1}, authors propose an iterative channel estimation method by introducing a trellis-based channel estimator for uplink NOMA.



\begin{figure*}[t] \vspace{0.3cm}
	\centering {\includegraphics[scale=0.9]{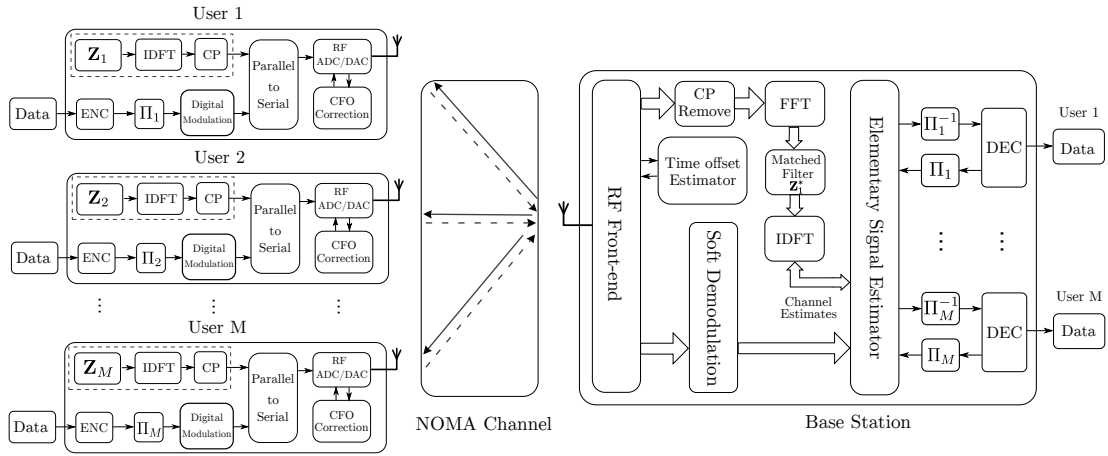}} 
	\caption{System model.}	 
	\label{sm}
\end{figure*}


Our proposed scheme is significantly different to \cite{noma_impl_1,noma_impl_2,noma_impl_3,noma_ce_ref1}, and our contributions can be summarized as follows. The uplink users estimate the CFO by using a reference signal transmitted by the BS and compensate it at the radio frequency (RF) front-end. The reference signal adopts the primary synchronization signal (PSS) of LTE.  We design a novel preamble for each uplink NOMA user, and then,  and it is appended to every frame before it is transmitted through the air. This preamble will be used for time synchronization, and channel estimation at the BS. Also, a low-complexity, iterative linear minimum mean squared error (LMMSE) detector has been implemented for multi-user decoding \cite{lei_lmmse}. Our proposed architecture is suitable for a frequency division duplex (FDD) uplink NOMA system, where the uplink frequency is different from the downlink, and hence channel reciprocity does not apply. The paper also validates the proposed architecture numerically, as well as experimentally through a software defined radio (SDR) implementation. Furthermore, experimental validation of uplink NOMA is not common in the literature, and can be highlighted as another major contribution of this paper.

The paper organization is as follows. The system architecture is discussed in Section \ref{sa}. Section \ref{dai} discusses the proposed time synchronization and channel estimation method. Then, in Section \ref{nr}, we validate the proposed scheme through simulations, and Section \ref{er} shows that the proposed methodology can be in fact implemented on hardware, and the experimental results are given to further validate our results. Section \ref{con} concludes the paper.

\section{System Architecture} \label{sa}

We consider an uplink NOMA system consisting of a base station (BS) and $ M $ users that simultaneously transmit information to the BS on the same frequency. As we have depicted in Fig. \ref{sm}, the data sequence of the $ m $-th user is encoded by a low rate encoder into a $ L $-length codeword. The use of the low rate encoder is to spread the data over a large bandwidth, since it guarantees a certain robustness against multipath propagation \cite{low_rate}. Then, the codeword is interleaved by the interleaver $ \Pi_m $ to reduce the effect of burst errors. Each user has its own unique interleaving pattern, and it is known a priori at the BS. Next, the interleaved sequence is mapped to a constellation, and then, it is modulated by a digital modulation scheme. Preamble is appended to the modulated signal before it is transmitted through the air. 

It is well known that CFO is  one of the most common impairments found in a practical communication system. Moreover, CFO leads to a large amount of multiple access interference in the uplink NOMA system, since CFO misaligns the sub-carriers in the frequency domain. Therefore, CFO estimation and compensation (CFO synchronization) becomes necessary. In practice, there are two common ways of CFO synchronization, \textit{i.e.}, uplink users estimate the CFO by using a reference signal transmitted by the BS and compensate it at the radio frequency (RF) front-end, or the BS estimates the CFO using the signals transmitted by uplink users and compensate it in the signal domain. For a uplink NOMA system, obviously the first method is easy to implement. Therefore, in our implementation, uplinks users perform CFO synchronization when BS transmits a reference signal.   

After CFO synchronization, it is necessary to ensure that the symbols transmitted by the superimposed users are time-aligned at the BS. Therefore, uplink users should arrange their timing advance  in such a way that all signals arrive to the BS simultaneously. In LTE, the BS continuously estimates the timing offset of  each UE and adjusts the uplink transmission timing by sending the value of timing advance to the respective uplink user. Also, it is well known that the channel estimation is a vital part in the receiver designs. However, in uplink NOMA, when the multiple users share the same time, frequency, and spreading code, estimating the timing offsets and channels  are not straightforward when compared to a non-NOMA system (e.g. LTE). Therefore, in this implementation, we design an unique preamble for each user $ m $, which can be used to estimate the time offsets and  the channels between the users and the BS in an uplink NOMA system.

We consider a low complexity iterative linear minimum mean squared error (LMMSE) detection \cite{lei_lmmse}. 
After the time synchronization and the channel estimation are done,
elementary signal estimator (ESE) calculates initial course estimates for the decoding process by jointly processing the signals of all users. We use the LMMSE estimator as the ESE. Also, the input to the ESE is the soft-demodulated output of the received composite signal, channel estimates, and a prior probability of occurrence of a symbol. 
ESE produces a soft outputs, which is given as input to the corresponding decoders. Decoder performs a extrinsic probability decoding in order to improve the estimates provide by the ESE. The ESE and decoders work in iterative manner to improve the results. Moreover, both ESE and decoders are based on soft in soft out principle. Each decoder handles data for a single user only, hence the complexity of a decoder is independent of the number of users. 


In the next section, we will discuss how the proposed preamble is designed, and how it can be used to estimate the time offsets and  the channels between the users and the BS in an uplink NOMA system.

\section{Channel estimation and Time Synchronization } \label{dai}

We will first present the proposed scheme for the channel estimation to draw useful insights, assuming there are $ M $ uplink users and perfect time synchronization, and then, 
we will extend the proposed scheme to the general case of having time misaligned users, where we want to estimate timing offset of each user.

As we have depicted in Fig. \ref{sm}, the symbol block  $ \mathrm  {\textbf{Z}}_m = [ Z_m(0),Z_m(1), \dots, Z_m(N-1) ] $ of the preamble  for the $ m-$th user is generated in the frequency domain (FD). Then, the discrete time (DT) signal $ \mathrm  {\textbf{z}}_m = [ z_m(0),z_m(1), \dots, z_m(N-1) ] $  is obtained by taking the $ N-$point inverse discrete Fourier transform (IDFT), where 
\begin{alignat*} {2}
z_m(n)
&=\displaystyle \frac{1} {\sqrt{N}}\sum_{k=0}^{N-1} Z_m(k) \exp\Big(j \frac{2\pi n k}{N} \Big).
\end{alignat*}
A cyclic prefix (CP), which is a copy of the last $ L  $ symbols of the block  $ \mathrm  {\textbf{z}}_m $, is pasted in front of the block $ \mathrm  {\textbf{z}}_m $, and the resulting signal can be written as
\begin{alignat*} {2}
\tilde z_m(n)
&= \begin{cases}
z_m(n+N-L),  & 0 \leq n <L \\
z_m(n-L),  & L \leq n<L+N-1
\end{cases} .
\end{alignat*}
Next, the DT signal $ \mathrm  {\tilde{\textbf{z}}}_m = [ \tilde z_m(0), \tilde z_m(1), \dots, \tilde z_m(N-1) ] $ is appended to each frame before it is transmitted through the air.

To this end, CP works as a guard interval between subsequent blocks, avoiding
the inter symbol interference (ISI) due to multipath fading. Therefore, if the length of the CP is longer than the maximum delay spread ($ L_p$ symbols) of the channel, \textit{i.e.}, $L > L_p$, or in other words, longer than the channel impulse response, ISI is prevented. Also, the use of the CP converts the DT linear convolution between the transmitted signal $ \tilde z_m(n) $ and the  $ m $-th user's channel impulse response $ \mathrm  {\textbf{h}}_m = [h_m(0),h_m(1), \dots, h_m(L_p-1) ] $ into a DT circular convolution. Hence, the received signal can be modelled as a circular convolution  between the channel impulse response and the transmitted data block \cite{cyclic_p}. Thus, assuming that uplink users have compensated the CFO, the received composite signal of the $ M $ users at the BS can be written as 
\begin{alignat*} {2}
\tilde y(n)
&=\displaystyle \sum_{m=1}^{M} \sum_{l=0}^{L_p-1} h_m(l) \tilde z_m(n-l) + v(n),
\end{alignat*}
where $ 0 \leq n < N+L+L_p-1 $, and $ v(n) $ represents the effect of noise. After the removal of the CP by taking the $ N $ samples starting from the $ L- $th sample point,  we have
\begin{alignat} {2}
y(n)
&= \tilde y(n+L) \nonumber \\
&=\displaystyle \sum_{m=1}^{M} \sum_{l=0}^{L_p-1} h_m(l) \tilde z_m(n+L-l) + v(n+L) \nonumber \\
&=\displaystyle \sum_{m=1}^{M} \sum_{l=0}^{L_p-1} h_m(l)  z_m(n-l) + v(n+L),
\end{alignat}
where $ 0 \leq n < N $. It should be noted that $ z_m(n) $ is periodic with period $ N $ due to the use of CP.

After applying the FD matched filter to $ y(n) $ with the first user's preamble sequence $ Z_1(k) $, the resultant signal can be written as
\begin{multline} 
G(k)
= \displaystyle \Bigg[ \frac{1} {\sqrt{N}}\sum_{n=0}^{N-1} y(n) \exp\Big(-j \frac{2\pi n k}{N} \Big) \Bigg] \odot \frac{1} {\sqrt{N}} Z_1^\ast(k) \\
\ \ \ \ \ \ \ \ \ \ = \displaystyle \Bigg[  {\sqrt{N}}\sum_{m=1}^{M}H_m(k) \odot Z_m(k) + V(k) \Bigg] \odot \frac{1} {\sqrt{N}} Z_1^\ast(k)  \\
=\displaystyle H_1(k)  +\sum_{m=2}^{M}H_m(k) \odot Z_m(k)  \odot  Z_1^\ast(k)  + \\ \hspace{4cm} \frac{1} {\sqrt{N}} V(k)  \odot Z_1^\ast(k) \\
\hspace{-1.2cm} = \displaystyle H_1(k)  +\sum_{m=2}^{M}H_m(k) \odot W_m(k)   + \\ \frac{1} {\sqrt{N}} V(k)  \odot Z_1^\ast(k), \label{matched_filter}
\end{multline} 
where $ \odot $ denotes element-wise multiplication, $ (\cdot) ^\ast$ denotes complex conjugate, 
\begin{alignat} {2}
H_m(k)
&= \frac{1} {\sqrt{N}}\sum_{l=0}^{L_p-1} h_m(l) \exp\Big(-j \frac{2\pi l k}{N} \Big) \nonumber \\
&= \frac{1} {\sqrt{N}}\sum_{n=0}^{N-1} \tilde h_m(n) \exp\Big(-j \frac{2\pi n k}{N} \Big), 
\end{alignat}
when
\begin{alignat} {2}
\tilde h_m(n)
&= \begin{cases}
h_m(n),  & 0 \leq n <L_p \\
0,  & L_p \leq n<N
\end{cases} , \label{ch}
\end{alignat}
and $ W_m(k) =  Z_m(k)  \odot  Z_1^\ast(k)$. It should be noted that $ \tilde h_m(n) $ only has non-zero values at the first $ L_p $ time indices. This is because $ L_p $ is the maximum delay spread of the channel ($ L_p $ is the length of channel impulse response). 
Now, by taking the $N-$point IDFT of \eqref{matched_filter}, we have 
\begin{alignat} {2}
g(n)
&= \frac{1} {\sqrt{N}}\sum_{k=0}^{N-1} G_1(k) \exp\Big(j \frac{2\pi n k}{N} \Big) \nonumber \\
&= \tilde h_1(n) + \sum_{m=2}^{M} \tilde h_m(n) \otimes w_m(n) + \tilde v(n) ,  \label{final_time_domain}
\end{alignat}
where $ \otimes $ denotes the cyclic convolution, and $ w_m(n)  $ is the $N-$point IDFT of $ W_m(k) $. 

\begin{figure}[t] \vspace{-0.0cm}
	\centering {\includegraphics[scale=0.55]{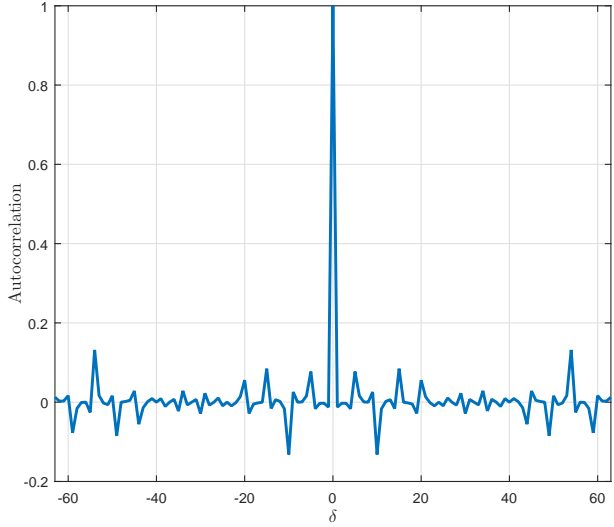}} 
	\caption{The cyclic autocorrelation of the Zadoff-Chu sequence.}	 
	\label{zc}
\end{figure}

In \eqref{final_time_domain}, there is a cyclic shift  on $ \tilde h_m(n)$ due to the cyclic convolution $ \tilde h_m(n) \otimes w_m(n)  $. To this end, if the cyclic shift  on $ \tilde h_m(n)$ is greater than $ L_p(m-1) $, $\brparen{\tilde h_m(n)}_{m=1}^{M}$  will not overlap each other. This is because $ \tilde h_m(n) $ only has non-zero values at the first $ L_p $ time indices. For example, when $ \mathrm  {\textbf{g}} = [ g(0),g(1), \dots, g(N-1) ] $ and the cyclic shift  on $ \tilde h_m(n)$ is  $ (L_p+1)(m-1) $, 
\begin{multline*} 
\mathrm  {\textbf{g}}
=   [ \tilde h_1(0),\cdots, \tilde h_1(L_p-1),0,\tilde h_2(0),\dots, \tilde h_2(L_p-1), 0, \cdots, \\ 0,  \tilde h_M(0), \cdots,\tilde h_M(L_p-1),0,\cdots,0 ]   \\
\hspace{-3cm} = [ \mathrm  {\textbf{h}}_1,0, \mathrm  {\textbf{h}}_2,0,\dots, \mathrm  {\textbf{h}}_M, 0, \cdots,0]. \hspace{3.1cm}
\end{multline*} 
Hence, if we want to recover the channel response of each user $ m $,  its preamble sequence should be the cyclic shifted version of the first user's preamble, with the cyclic shift $ L_p(m-1) $. According to the $ N-$point DFT property,  $ W_m(k) =  Z_m(k)  \odot  Z_1^\ast(k) =  \exp\Big(j \frac{2\pi L_p (m-1) k}{N} \Big)$, \textit{i.e.},
\begin{alignat} {2}
Z_m(k)
&= Z_1(k) \exp\Big(j \frac{2\pi L_p (m-1) k}{N} \Big).  
\end{alignat}
Furthermore, we use so-called Zadoff-Chu sequence as the first user's preamble sequence $ Z_1(k) $, $ i.e. $,
\begin{alignat*} {2}
Z_1(k)
&= \begin{cases}
\exp\Big(-j \frac{\pi \gamma k^2}{N} \Big),  & N \ \ \text{even} \\
\exp\Big(-j \frac{\pi \gamma k(k+1)}{N} \Big),  & N \ \ \text{odd}
\end{cases} ,
\end{alignat*}
where $ N $ is the length of the sequence, $ \gamma $ is the root index relatively prime to $ N $, and $ k \in \{1, \dots, N-1 \} $. The values of $ \gamma $ and $ N $ are same for all users belong to the BS. This sequence exhibits a useful property that the cyclically shifted versions of themselves are orthogonal to one another. In addition to that, the cyclic autocorrelation of each sequence results in a single Dirac-impulse at time offset zero \cite{Zchu}, see Fig. \ref{zc}. Therefore, this property allows us to estimate the timing offsets  when the uplink users are time misaligned.


In this implementation, we use a maximum likelihood (ML) detector to find the timing offset $ \delta^\star_m $ of the $ m-$th user. Moreover, ML detector finds $ \delta^\star_m $ that corresponds to the position of the maximum in the correlation between the received signal $ y(n)  $ and the preamble sequence $ z_m(n) $, \textit{i.e.},
\begin{equation} 
\delta^\star_m =  \argmax_\delta \left \{ \sum_{n=0}^{N-1} y(n+\delta)z_m^\ast(n) \right \}.
\end{equation}
We should stress that the time offset estimation was mainly possible due to the selection of Zadoff-Chu sequence as the first user's preamble sequence. BS continuously estimates $ \delta^\star_m $ and adjusts the uplink transmission timing by sending it to the
respective uplink user. In the next section, we will validate our results using numerical evaluations.

%
%
%
%
%
%

\section{Numerical Results} \label{nr}

\begin{figure}[t] \vspace{-0.0cm}
	\centering {\includegraphics[scale=0.52]{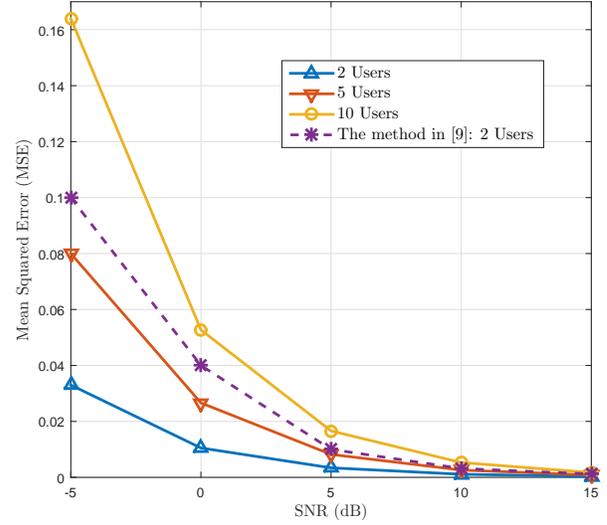}} 
	\caption{The behaviour of the channel estimation error for different SNR values.}	 
	\label{cse}
\end{figure}

In this section, we present some numerical examples to validate our proposed schemes, and to provide useful insights on uplink NOMA channel learning and low-complexity iterative
LMMSE detector. Fig. \ref{cse} illustrates the behaviour of the channel estimation error for different SNR values. As expected, for higher SNR values, we have lower mean squared error (MSE) values. The important point to notice in the figure is that the MSE has increased with number of users. However, the loss is rather acceptable given the practicality of the proposed method. Furthermore, figure illustrates that our proposed method allows to achieve significant gains when compared to other works in the literature, even with lower SNR values.

\begin{figure}[t] \vspace{-0.0cm} 
	\centering {\includegraphics[scale=0.52]{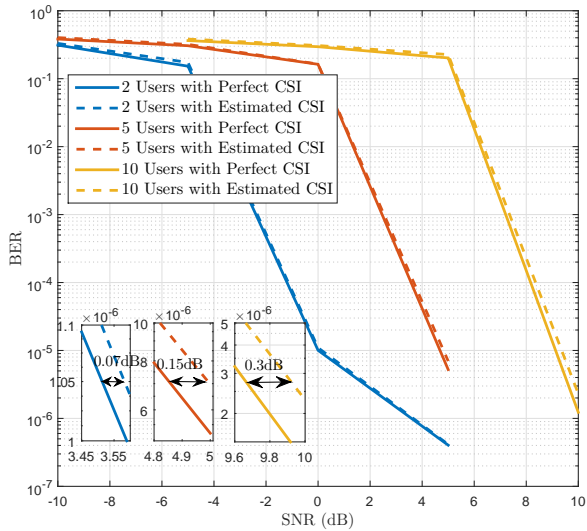}} 
	\caption{The behaviour of the BER curves.}	 
	\label{ber}
\end{figure}

Fig. \ref{ber} illustrates the bit error rate (BER) curves of the proposed uplink NOMA scheme. We used the LMMSE estimator as the ESE and the channel estimation was done using our proposed scheme. We have simulated the repetition-aided irregular repeat-accumulate (Rep-IRA) code \cite{rep_code} over NOMA. The Rep-IRA code is  constructed by parallelly  concatenating a rate-$1/5$ repetition code and a non-systematic IRA code. The degree sequence of the  IRA code is ${{\lambda}}(x)=0.171579x+0.284322x^2+0.030637x^9+0.513463x^{29}$. The length and the code rate of the Rep-IRA code is $4096$ and $0.1$, respectively. In the simulation, we assume that each user employs a random interleaver and a quadrature phase-shift keying (QPSK) modulator. We have given two BER curves for each user scenario, \textit{i.e.}, $ M=2,M=5, $ and $ M=10 $, when the BS estimates the channels using our proposed method, and when the BS has perfect CSI. The figure shows that the  gap between two BER curves (perfect CSI curve and imperfect CSI curve) increases with the increase of number of users. However, the BER gaps are not significantly large. Having done the numerical evaluations, we will further validate our proposed architecture experimentally in the next section.

\section{experimental validation} \label{er}

\begin{figure}[t]  \vspace{0.0cm}
	\centering {\includegraphics[scale=0.8]{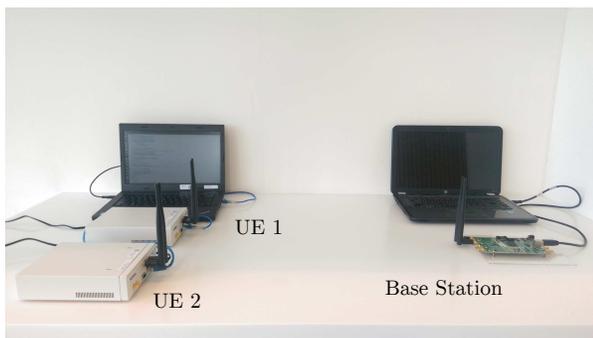}} 
	\caption{Experimental setup.}	 
	\label{es} \vspace{-0.3cm}
\end{figure} 

Our experimental setup consists of  a BS and two users, see Fig. \ref{es}. The two uplink users and the BS have been implemented using SDRs in conjunction with laptops having Ubuntu 14.04. The other system parameters used in our implementation is summarized in Table I.  All the real-time signal processing tasks, \textit{i.e.}, CFO synchronization, time synchronization, channel estimation, modulation, soft-demodulation,  were performed on  laptops using the GNU Radio framework, an open-source software development tool-kit which facilitates us to implement signal processing modules for SDRs. USRP hardware driver (UHD) is the interface between GNU Radio and SDR. However, our current uplink NOMA system has not been sufficiently integrated, BS can not operate with all the necessary functions in real time for the iterative decoding process. Therefore, offline experiments has been carried out for evaluating the over-the-air performance of uplink NOMA, where the  soft-demodulated data and the channel estimates are dumped to a file by the BS, and then, which are used to recover the user transmit data using a simulation program.

 
\begin{figure}[t]  \vspace{0.54cm}
	\centering {\includegraphics[scale=0.8]{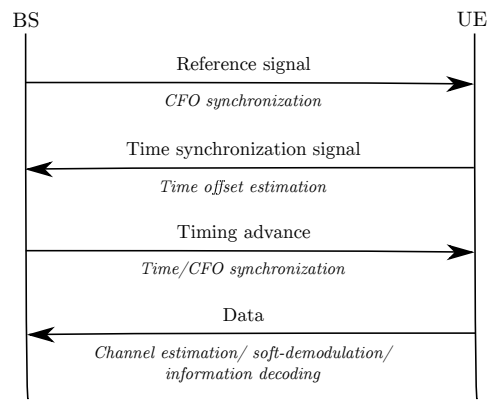}} 
	\caption{Uplink NOMA transmission procedure.}	 
	\label{pro}
\end{figure}

\begin{table} 
	\caption {System configuration.} 
	\rowcolors{1}{}{lightgray}
	\centering \begin{tabular}{  p{3cm}  p{4.5cm}  }
		\hline  \centering Transmission mode   &  \centering\arraybackslash SISO \\ 
		 \hline \centering BS (SDR) & \centering\arraybackslash USRP B210  \\
		 \hline  \centering UEs (SDR) & \centering\arraybackslash USRP 2920 \\
		 \hline  \centering USRP hardware driver (UHD)  & \centering\arraybackslash 3.8 \\
		 \hline  \centering Channel estimation \& Time synchronization & \centering\arraybackslash Cyclically shifted preambles \\
		 \hline  \centering CFO synchronization & \centering\arraybackslash Primary Synchronization Signal (LTE)  \\
		  \hline  \centering Channel coding  & \centering\arraybackslash Turbo Hadamard \\
		  \hline  \centering Carrier frequency  & \centering\arraybackslash 915MHz \\
		  \hline  \centering Detector  & \centering\arraybackslash LMMSE \\
		 \hline  \centering Modulation scheme &  \centering\arraybackslash SC-FDMA with BPSK symbols \\ \hline
		
	\end{tabular}

\end{table} 

Our intended uplink NOMA transmission procedure is given in Fig. \ref{pro}. Firstly, the BS broadcasts the reference signal for CFO synchronization. The reference signal  adopts the primary synchronization signal (PPS) of LTE. To this end, each UE estimates the CFO and compensates it by adjusting its local oscillator frequency. Next,  UEs transmit the time synchronization signal, which has been generated by appending the preamble to a dummy data sequence. The BS estimates the timing offset of  each UE and adjusts the uplink transmission timing by sending the value of timing advance to the respective uplink user. This time synchronization step can be repeated several times until the symbols transmitted by the superimposed users are time-aligned at the BS. Finally, UEs start the uplink transmission, BS performs channel estimation, soft-demodulation, and iterative information decoding. 

\begin{figure}[t]  \vspace{0.2cm}
	\centering {\includegraphics[scale=0.8]{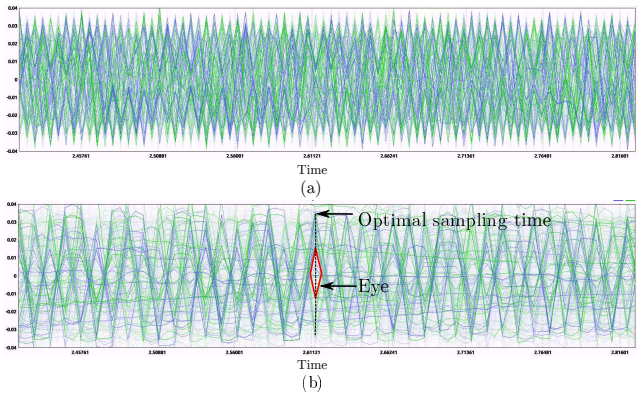}} 
	\caption{Eye diagrams.}	 
	\label{eye}
\end{figure}

Fig. \ref{eye} illustrates the eye diagram,  which is the all possible realisations of the received signal of interest viewed within a particular signalling interval. An open eye pattern corresponds to  minimal signal distortion. To this end, Fig. \ref{eye}(a) is the eye diagram when the two users are not well synchronized (time/CFO). The closure of the eye pattern is evident that there is a high distortion in the received signal. We can see an open eye pattern, when we use our proposed uplink NOMA scheme, see Fig. \ref{eye}(b). The optimal sampling time of a symbol is the midpoint of an eye. Also, having the optimal sampling time allows us to  do the soft-demodulation accurately, and it increases the system performance. Furthermore, we calculated BER values (0.1 $ \sim $ 0.001) through our  extensive experiments. These BER values are rather acceptable given the practicality of the proposed system.

\section{Conclusions} \label{con}

This paper has presented an over-the-air implementation of an uplink NOMA system, which is not common in the literature. We have discussed most common practical difficulties, when it comes to a real world uplink NOMA implementation,  while providing solutions to most important problems, \textit{i.e.}, carrier frequency offset (CFO) synchronization, time synchronization, and channel estimation. The implemented CFO synchronization method adopts the primary synchronization signal of LTE. We proposed a novel preamble architecture for uplink NOMA users, and it is used for time synchronization and channel estimation of the system. Also, a low-complexity, iterative LMMSE detector was implemented for multi-user decoding. The paper also validates the proposed architecture numerically, as well as experimentally through a SDR implementation. Further analytical and experimental results related to the proposed architecture will be presented in future work.


\bibliography{bibfile}
\end{document}